# Particle–fluid interactivity deteriorates buoyancy driven thermal transport in nanosuspensions: A multi component lattice Boltzmann approach

**Savithiri S, Purbarun Dhar [#], Arvind Pattamatta and Sarit K. Das\***

Department of Mechanical Engineering, Indian Institute of Technology Madras,

Chennai–600036, India

[#] E–mail: pdhar1990@gmail.com

*Corresponding author: E–mail: skdas@iitm.ac.in

## Abstract

Severe contradictions exist between experimental observations and computational predictions regarding natural convective thermal transport in nanosuspensions. The approach treating nanosuspensions as homogeneous fluids in computations has been pin pointed as the major contributor to such contradictions. To fill the void, inter–particle and particle–fluid interactivities (slip mechanisms), in addition to effective thermophysical properties, have been incorporated within the present formulation. Through thorough scaling analysis, the dominant slip mechanisms have been identified. A Multi Component Lattice Boltzmann Model (MCLBM) approach has been proposed, wherein the suspension has been treated as a non–homogeneous twin component mixture with the governing slip mechanisms incorporated. The computations based on the mathematical model can accurately predict and quantify natural convection thermal transport in nanosuspensions. The role of slip mechanisms such as Brownian diffusion, thermophoresis, drag, Saffman lift, Magnus effect, particle rotation and gravitational effects have been pictured articulately. A comprehensive study on the effects of Rayleigh number, particle size and concentration reveals that the drag force experienced by the particles is dominantly responsible for deterioration of natural convective thermal transport. In essence, the dominance of Stokesian mechanics in such thermofluidic systems is established in the present study. For the first time, as revealed though thorough survey of existent literature, a numerical formulation explains the contradictions observed, rectifies the approach, predicts accurately and reveals the crucial mechanisms and physics of buoyancy driven thermal transport in nanosuspensions.

# 1. Introduction

Dilute suspensions of nanostructures in fluid media; dubbed as 'nanofluids' in scientific colloquialism, have been a topic of tremendous interest among researchers and developers in the academic community. The interest spawns as a result of the enhanced thermal transport phenomena such suspensions exhibit as working medium in thermofluidic systems compared to the base fluids. Nanosuspensions have been widely reported to exhibit augmented thermal conductivities [1-5] and forced convective heat transfer coefficients [6-10], as obtained from experiments as well as through computational modeling. In case of forced convection, the numerical predictions, even those utilizing effective property models, accurately predict the trend, i.e. enhanced transport as obtained from experiments. However, this is not so in case of natural convective heat transfer in such nanosuspensions. Experiments unanimously reveal that heat transfer due to thermally induced buoyancy driven natural convection deteriorates for nanosuspensions as compared to the caliber of the base fluid [11-15] the deterioration being augmented with increasing concentration of the dispersed phase. Au contraire, reports on numerical simulations of natural convective thermal transport in nanosuspensions always predict the exact opposite.

Mathematical models for convective thermal transport in nanosuspensions proposed so far by researchers can be broadly classified by three methods. The first being the single component model; wherein the suspension is considered to be a single fluid with effective properties by assuming distribution of nanoparticles within the fluidic matrix to be homogeneous [16-19]. The method has been observed to predict enhanced natural convective thermal transport. The second method involves a single component nonhomogeneous model [20-24], wherein the suspension is considered as a single fluid incorporating particle distribution, i.e. a pseudo-single fluid model. This method has also resulted in inaccurate predictions. The final model encompasses the particle distribution and segregates it from the fluidic phase, thereby creating a multi-component model. In recent times, a few reports on usage of the multi-component system to predict transport in colloidal media have surfaced within the academic community; however, the simulations could not quantify the deteriorated transport in natural convection of nanosuspensions nor the cause of such deteriorations. Lattice Boltzmann simulations to simulate flow and thermal transport in nanosuspensions utilizing the multi component model [25] reveals that it can effectively track the concentration distribution as time evolves and that Brownian motion plays an important role as a slip mechanism, both of which are essential in governing heat transport process. Similar usage of the model for predicting convective heat transfer in nanosuspensions in a square cavity has also been reported [26], wherein it has been concluded that the critical Rayleigh number (Ra) in case of the suspension is lower than the base fluid and enhanced Ra leads to enhanced heat transfer due to thinning of the boundary layer adjacent to the vertical walls.

Furthermore, two-phase Lattice Boltzmann models (LBM) [27] incorporating the interaction forces of Brownian, gravity, drag and interaction potential, and utilized to solve natural convection in nanosuspensions lead to counter-observed results. It predicted enhancement in average Nusselt number with increase in particle concentration and Ra. However, such findings are exactly antiparallel to the experimental observations. As a result, a LBM formulation treating the particle and the fluidic phase separately, yet solving them as a coupled system is essential to bridge the gap between experimental and computational findings. The present paper determines drag force as the most dominant force as far as natural convection is concerned, through scaling analysis. Based on the theoretical analysis, a complete multi component mathematical model is developed so as to segregate the fluidic and particulate phases, while in essence solving them as a coupled system. The system of equations are solved for a domain simulating experimental reports utilizing a LBM formulation, so as to specifically incorporate the inter-particle and particle-fluid interactivities within the scope of the simulation. Comparison of obtained transport parameters with experimental reports yield that the present approach accurately; both qualitatively and quantitatively, predicts the deteriorated transport. A detailed parametric study so as to understand the role of slip mechanisms and material constraints on the deterioration of heat transfer has been performed and the underlying physics of buoyancy driven thermal transport in nanosuspensions has been conclusively brought forward.

## 2.  Mathematical Formulation

## 2. a. Governing Transport Equations

In order to implement the MCLBM formulation to the problem of natural convection thermal transport in nanosuspensions, recognizing the proper form of the transport equations is necessary, which in turn requires comprehending the proper and allowable simplifying assumptions.

The present formulation assumes a continuum model for the nanosuspensions, such that the Knudsen number, Kn < 0.3. The assumption is theoretically backed by the fact that the inter–particle interactions which are mapped by the MCLBM formulation occur across mean paths of magnitudes greater than the particle diameter. Furthermore, if particle–fluid molecule interactions are considered, even then the ratio of fluid molecule diameter to particle diameter falls much short of unity, and as such, the largest conceivable interaction based Kn that needs to be accounted for still falls within the continuum regime.

The slip mechanisms incorporated in the formulation have been verified to have a dominant control over the natural convection phenomena from detailed scaling analysis [28], viz. Brownian diffusion, thermophoresis, drag, Saffman's lift, Magnus effect, particle rotation and gravitational effects and the need for such particle migratory effects and the corresponding physics has been discussed in details in the next section. Only dilute nanosuspensions have been considered in the present study, as thereby the interparticle slip mechanisms of van der Waal's forces, particle agglomeration de–agglomeration kinetics and electrostatic interactions caused by the interfacial electric double layer; which are essentially features of concentrated systems; have been assumed to have negligible effects as compared to the seven mechanisms discussed. Furthermore, the forces generated due to the slip mechanisms are distributed randomly in spatial coordinates; i.e. they are essentially vectors in a three dimensional coordinate system. However, since the present formulation is restricted to 2D space the variations are confined within the domain plane. Therefore, the forces can be assumed to be scalars for sake of computational simplicity and the planar directionality is accounted for by the LBM formulation.

Having stated the assumptions behind formulating the governing equations for the system, the latter can now be defined. Continuity (Eq. 1), momentum (Eq. 2) and energy (Eq. 3) transport equations for the nanosuspension and equation of motion for the nanoparticles (Eq. 5) are solved to reveal system behavior during natural convection thermal transport. The equations are expressible in terms of the nanosuspension velocity vector '$u$' and temperature '$T$' as follows

$$\nabla . u = 0 \tag{1}$$

$$\rho_{ns}\left[\frac{\partial u}{\partial t} + u.(\nabla u)\right] = -(\nabla p) + \nabla\left[\mu_{ns}(\nabla u + \nabla u^T)\right] + \rho_{ns}\beta_{ns}\Delta T - S_{np} \tag{2}$$

$$\rho_{ns} C_{p,ns} \left[ \frac{\partial T}{\partial t} + u.\nabla T \right] = \nabla(k_{ns}.\nabla T) \tag{3}$$

where, '$\rho$', '$\mu$', '$\beta$', '$C_p$' and '$k$' represent density, viscosity, specific heat, coefficient of thermal expansion and thermal conductivity. Subscript '$ns$', '$p$' and '$f$' refers to the nanosuspension, nanoparticles and the base fluid respectively whereas superscript '$T$' refers to transpose operator. '$S_p$' is the source term incorporated in order to accommodate the slip mechanisms of the nanoparticles and the consequent migration within the suspension. It is expressed as the net slip force '$F$' per unit volume '$V$' of the particle as

$$S_p = \frac{\sum F}{V_p} \tag{4}$$

The net force is utilized to determine the motion of the nanoparticle under the influence of the seven slip mechanisms. In the Lagrangian frame of reference, the equation for nanoparticle motion can be expressed as

$$m_p \frac{dv_p}{dt} = \sum F \tag{5}$$

The net force may be expressed as the algebraic sum of the forces generated by the seven slip mechanisms as

$$\sum F = F_B + F_T + F_D + F_M + F_R + F_S + F_G \tag{6}$$

where, subscripts $B$, $T$, $D$, $M$, $R$, $S$ and $G$ represent Brownian diffusion, thermophoresis, drag, Magnus, Saffman lift and gravitational respectively. The slip mechanisms and the corresponding force functions have been discussed in details in the next section.

Particle flux independent properties such as the density, specific heat and coefficient of thermal expansion are deduced from the concentration '$\varphi$' based on the classical Maxwell–Garnett effective medium models as

$$\rho_{ns} = \phi \rho_p + (1-\phi)\rho_f \tag{7}$$

$$(\rho C_p)_{ns} = \phi(\rho C_p)_p + (1-\phi)(\rho C_p)_f \tag{8}$$

$$\beta_{ns} = \phi\beta_p + (1-\phi)\beta_f \tag{9}$$

Properties of the nanosuspension which are dependent on dynamic factors (concentration and temperature induced particle migration), such as thermal conductivity and viscosity are determined based on pertinent mathematical models and experimental results [3, 5, 14]. While the thermal conductivity is governed by the migratory effects of the particles due to thermal fluctuations, the viscous nature has been considered to follow similar trend as the base fluid with respect to temperature. They are expressible in terms of the temperature dependent Prandtl number '$Pr(T)$' and particle migration Reynolds number '$Re(T)$' as [29]

$$\frac{k_{ns}}{k_f} = 1 + 64.7\phi^{0.764}\left(\frac{d_f}{d_p}\right)^{0.369}\left(\frac{k_f}{k_p}\right)^{0.7476} \Pr(T)^{0.995} \Re(T)^{1.2321} \tag{10a}$$

$$\Pr(T) = \frac{\mu_f}{\rho_f \alpha_f}, \quad \Re(T) = \frac{\rho_f K_B T}{3\pi\mu_f^2 d_f} \tag{10b}$$

$$\frac{\mu_{ns}(T)}{\mu_f(T)} = 1 + 4.93\phi + 222.4\phi^2 \tag{11}$$

Based on the above mentioned equations (1) to (11) and a thorough scaling analysis of the slip mechanisms to be discussed in the next section, the order of magnitude and degree of influence on each mechanism on the buoyancy driven transport is determined. Furthermore, the system of transport equations is extended to determine physics of natural convective thermal transport in nanosuspensions in a square cavity and validated with experimental observations.

## 2. b. Slip Mechanisms: Force Functions and Time Scales

Nanosuspensions have been widely reported to exhibit augmented thermal conductivity and viscous characteristics [14, 30] compared to the dispersing medium. The marked enhancement observed in thermal conductivity of the system is often alleged to be the prime source of deterioration of convective transport characteristics via the Nusselt number formulation. This however drastically fails to quantitatively predict convective heat transfer characteristics, both forced, wherein an enhancement is observed, as well as natural, where the degree of deterioration is much more pronounced than that obtained by the Nusselt number approach. It is plausible to put forward an argument for reduction in natural convection heat transfer based on the enhanced viscosity of the suspension. Augmented viscous behavior leads to enhanced momentum diffusivity within the buoyancy driven flow, leading to higher flow damping and thereby reduction in thermal transport. However, this too cannot predict the degree of deterioration. Particle migration within the colloidal phase has often been regarded as a major source of enhanced force convection thermal transport. Such an approach has been extended to the buoyancy driven transport and inter–particle forces and particle–fluid interactivities should be incorporated within the governing formulation so as to evolve the proper transport equations. The particle–fluid slip mechanisms lead to the generation of corresponding rate of change of momentum between the fluid and the particles which need to be incorporated into the momentum transport equations so as to obtain the correct form of governing equations for the twin component system. The expressions for the seven slip generated force components, as described by Eqn. (6) have been tabulated in Table 1.

| Slip Mechanism | Expression for force function |
|---|---|
| Brownian diffusion | $F_B = \rho_p D_B \nabla \phi A_p v_B$, where, $D_B = \dfrac{2k_B T}{3\pi \mu_{ns} d_p}$ and $v_B = \dfrac{2k_B T}{\pi \mu_{ns} d_p^2}$ |
| Thermophoresis | $F_T = \rho_p v_T \phi A_p$, where, $v_T = \left[ \dfrac{6\pi C_S \{k^* + C_t(Kn)\}}{\{1+3C_m(Kn)\}\{1+2k^* + 2C_t(Kn)\}} \right]\left(\dfrac{\mu_{ns}}{\rho_{ns}}\right)\left(\dfrac{\nabla T}{T}\right)$ $C_S = 1.17$, $C_t = 2.18$, $C_m = 1.14$, $k^* = k_{ns}/k_p$ and $Kn = \lambda/d_p$ |
| Drag | $F_D = -\beta(v_f - v_p)m_p$, where, $\beta = \left(\dfrac{18\mu_{ns} C_d \operatorname{Re}_d}{24\rho_p}\right)$ $\operatorname{Re}_d = \left(\dfrac{\rho_p d_p |v_f - v_p|}{\mu_{ns}}\right)$ and $C_d = \dfrac{24}{\operatorname{Re}_d}\left(1+0.15\operatorname{Re}_d^{0.687}\right)$ |
| Saffman Lift | $F_S = \dfrac{K_S}{4} \mu_{ns} |v_f - v_p| d_P^2 \left(\dfrac{\rho_{ns}\dot{\gamma}}{\mu_{ns}}\right)$, where, $K_S = 81.2$ and $\dot{\gamma} = \dfrac{8v_p}{H}$ |

| | | |
|---|---|---|
| Magnus Effect | $F_M = \dot{m}_p(v_p - v_M)$, where, $v_M = 0.34\left(\dfrac{v_M^2 H}{v_{ns}}\right)\left(\dfrac{d_p}{H}\right)^{2.84}\left(\dfrac{2z}{H}\right)\left(0.63 - \dfrac{2z}{H}\right)$ | |
| | and $v_M = \dfrac{v_f + v_p}{2}$ | |
| Particle rotation | $F_R = \dfrac{\mu_{ns} \rho_p \dot{\gamma}}{\rho_{ns}}$ | |
| Gravitational effects | $F_G = -V_p(\rho_p - \rho_f)g$ | |

**Table 1**: Expression for the force functions of each dominant slip mechanism

where, '$v$', '$A$', '$\lambda$', '$C_d$', '$K_s$', '$\dot{\gamma}$' and '$H$' represents velocity, surface area, mean free path of particle migration, migration induced coefficient of Stokesian drag, coefficient of Saffman lift, shear rate and characteristic length scale for domain undergoing convective transport respectively. Likewise, subscripts '$p$', '$f$' and '$ns$' represent the particulate phase, base fluid and nanosuspension respectively.

Determining the extent to which each slip mechanism contributes to the deterioration of natural convective thermal transport, it is necessary to scale the Reynolds number of the particle as a function of all the Reynolds numbers borne out of each slip mechanism. This is possible by identifying the characteristic time scale for particle migration under the influence of each slip mechanism. The characteristic time scale '$\tau$', is defined as the mean time taken by the nanoparticles to diffuse across a distance equivalent to the diameter of the particle. The expression for '$\tau$' for each mechanism discussed has been tabulated in Table 2.

| Particle | Brownian | Thermo- | Drag | Saffman | Magnus | Rotation | Gravity |
|---|---|---|---|---|---|---|---|

|  |  | phoresis |  | Lift |  |  | Effect |
| --- | --- | --- | --- | --- | --- | --- | --- |
| $\tau_P$ | $\tau_B$ | $\tau_T$ | $\tau_D$ | $\tau_S$ | $\tau_M$ | $\tau_R$ | $\tau_G$ |
| $\dfrac{d_p}{v_p}$ | $\dfrac{d_p^2}{D_B}$ | $\dfrac{d_p}{v_T}$ | $\dfrac{d_p}{\lvert v_f - v_p \rvert}$ | $\dfrac{\rho_p d_p^2}{\mu_{ns}}$ | $\dfrac{d_p}{v_M}$ | $\dfrac{\rho_{ns} d_p^2}{\mu_{ns}}$ | $\dfrac{d_p}{v_G}$ |

**Table 2**: Expression for the characteristic time scale of each dominant slip mechanism

## 2. c. Scaling Analysis

Prior to implementation of the LBM formulation to any specific domain undergoing transport phenomena, it is important to utilize a simple scaling methodology performed on the vertical velocity so as to identify the degree of dominance that each of the scaling mechanism poses towards the process. Based on formulating the properly scaled expression for the particle Reynolds number under the combined influence of all the slip mechanisms, the order of magnitude of contribution of each mechanism towards deterioration of the transport process can be deduced. The analysis essentially involves comprehensive approach towards segregating the major diffusive terms in the governing equation and thereby treats the suspension as a single component fluid with localized diffusive fluxes. This analysis therefore provides insight into the individual capabilities of the slip forces and not a quantitative picture as the proposed MCLBM. However, it is important to understand the transport physics of the system. The order of magnitude expression for the particle Reynolds number is expressed as [28]

$$\begin{aligned}
\mathrm{Re}_p \sim &-18\left(\frac{\tau_P}{\tau_D}\right)(1+0.15\,\mathrm{Re}_D^{0.687}) - \mathrm{Re}_G\left(\frac{\tau_P}{\tau_G}\right)\left(\frac{d_p g}{v_G^2}\right) + \mathrm{Re}_B\left(\frac{\tau_P}{\tau_B}\right) d_p \nabla \phi \\
&+ \mathrm{Re}_T\left(\frac{\tau_P}{\tau_T}\right)\phi + \mathrm{Re}_L\left(\frac{\tau_P}{\tau_L}\right)\left(\frac{K_L d_p}{4\pi}\right)\left(\sqrt{\frac{\dot{\gamma}}{v_{ns}}}\right) + \mathrm{Re}_R\left(\frac{\tau_P}{\tau_R}\right) + \mathrm{Re}_M\left(\frac{\tau_P}{\tau_M}\right)
\end{aligned} \quad (12)$$

The order of magnitude of the particle mean velocity is calculated based on the particle Reynolds number defined in Eqn. 12 from definition as

$$\mathrm{Re}_P \sim \frac{\rho_P v_p d_p}{\mu_{ns}} \tag{13}$$

The order of magnitude of the local nanosuspension velocity is deduced from the conservation of volume flux of the fluid and the slip flux of the particles, per unit area of the domain as

$$(\rho v)_{ns} \sim \phi(\rho v)_p + (1-\phi)(\rho v)_f \tag{14}$$

The vertical convective transport velocities of the base fluid and the suspension can be deduced from scaling the thermal gradient induced buoyant body force component [31] as

$$v_f \sim \sqrt{(g\beta)_f \Delta Th} \tag{15a}$$

$$v_{ns} \sim \sqrt{(g\beta)_{ns} \Delta Th} \tag{15b}$$

The complete analysis described in the preceding segment enables deduction of the order of magnitude of the thermally driven buoyant component '$(g\beta)_{ns}$' which is further utilized to determine the order of the governing Rayleigh number for the suspension as a single component system. The Rayleigh number, expressed in Eqn. (16) incorporates the scaled effect of all the slip forces and is used to evaluate the thermal performance of the suspension from the Nusselt number expressed in Eqn. (17) [31].

$$Ra_{ns} = \frac{(g\beta)_{ns} \Delta Th^3}{(\nu\alpha)_{ns}} \tag{16}$$

$$Nu_{ns} = 0.18 \left[ \frac{(Ra\,\mathrm{Pr})_{ns}}{0.2 + \mathrm{Pr}_{ns}} \right]^{0.29} \tag{17}$$

From the analysis mentioned, it is possible to qualitatively deduce the deterioration in natural convection heat transfer caused due to the migratory motion of nanoparticles. The Nusselt numbers obtained from the scaling analysis have been plotted as a function of operating Rayleigh numbers and particle concentration and illustrated in Fig. 1 (a)–(d). It is evident from the scaling analysis itself that natural convective thermal transport in nanosuspensions is expected to exhibit deterioration compared to normal fluids and the deterioration is enhanced with increasing particle population. Furthermore, it can be observed that non-consideration of all other slip mechanisms other than drag leads to very small observational differences in the Nusselt number as compared to the case with all the slip mechanisms. However, when solely the drag force is not considered, an appreciable difference in thermal transport is observed and the deterioration is seen to be not as pronounced as the case when all the slip mechanisms are considered. This essentially indicates that in buoyancy driven transport of granular systems, drag force holds a position of major importance over all other slips in governing thermal transport. However, in order to quantify the same, a MCLBM formulation is required since in a Single Component Non-homogeneous Model system (SCNHM), which involves single component, the sole protocol to define fluidic drag acting on the particulate phase is to define a pseudo-diffusive term that essentially accounts for the drag force. Such a methodology however has been noted to fail in terms of quantitative perspectives, similar to the results of the scaling analysis.

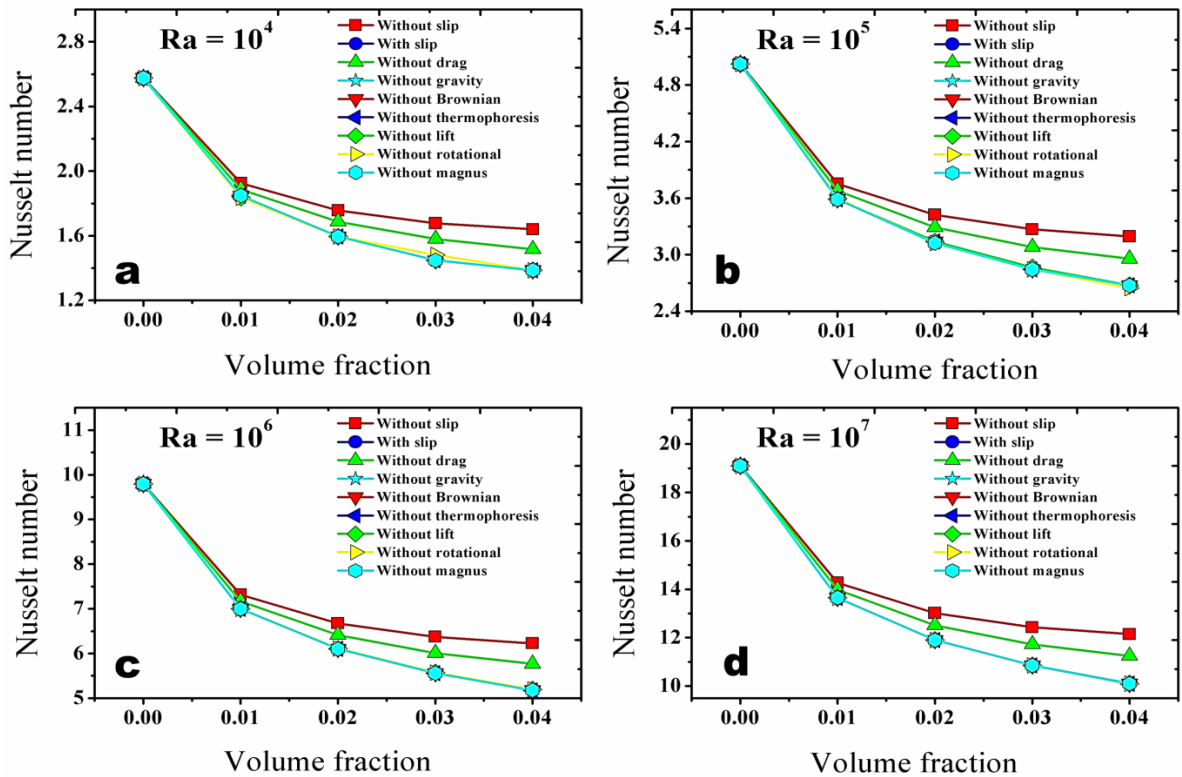

**Figure 1:** Natural convective heat transfer characteristics as obtained from scaling analysis

Fig. 1 also qualitatively reveals the effect of slip mechanisms combined with the strength of the thermally induced buoyant system on the thermal transport characteristics. The governing Rayleigh number seems to pose a strong influence on the ability of the slip forces to deteriorate the thermal transport. At Ra of the orders of $10^4$ and $10^5$, drag is observed to be more important followed by all other forms of slip, which appear to possess a quantitatively similar magnitude. However, as Ra increases to the order of $10^6$ and $10^7$, the deterioration caused by drag is largely augmented over all the other slip forces. For e.g., when a 4 % nanosuspension is considered, deterioration predicted by simplistic effective property analysis considering the non dimensionalized thermal gradient definition of Nusselt number is ~ 36% compared to the base fluid. When all slip mechanisms are included, this deterioration is around ~ 47% whereas when only drag is not considered, the deterioration is predicted as ~ 41%. This essentially signifies that the relative motion between bulk fluid motion and particle velocity strongly influences thermal transport deterioration, the mechanisms behind which is further elaborated in the Results and Discussion section.

## 2. d. Multi-Component Non-Homogeneous Formulation

The inadequacy of the scaling analysis to capture the magnitudes of deterioration in heat transfer as experimentally recorded occurs due to the order-of-magnitude approach towards the slip forces. In order to numerically simulate the actual deterioration the nanosuspensions need to be effectively modeled mathematically such that the fluid and particle phases are computationally separate coupled systems with the slip forces acting as the spatio-temporal coupling parameters. In order to achieve the same, a Lattice Boltzmann formulation has been developed which can essentially track the interactivities of the particulate phase with the fluid phase, thereby computationally segregating the two transport systems in order to accommodate the slip mechanisms easily as source functions. The LBM assumes the particulate and fluid phases to be mesoscopically located at the nodes of a spatial lattice structure [25]. Although the method cannot track singular particles in the suspension, it reduces the computational time and complexity by taking the ensemble of complicated forces and interactions involved in the nanosuspension into account. The BGK collision model approximated Lattice Boltzmann equation [32] for hydrodynamic (Eqn. 18) and thermal transport (Eqn. 19) for the fluid and the particulate phase can be expressed as

$$f_i^\sigma(x+e_i\Delta t, t+\Delta t) - f_i^\sigma(x,t) = -\frac{1}{\tau_{mom}^\sigma}(f_i^\sigma(x,t) - f_i^{\sigma,eq}(x,t)) + \Delta t W_i(F_i^\sigma e_i) \tag{18}$$

$$g_i^\sigma(x+e_i\Delta t, t+\Delta t) - g_i^\sigma(x,t) = -\frac{1}{\tau_\theta^\sigma}(g_i^\sigma(x,t) - g_i^{\sigma,eq}(x,t)) \tag{19}$$

In the preceding equations, $f_i$ and $g_i$ represent particle density and temperature distribution functions for the finite set of particle velocity vectors represented by $e_i$. The superscript 'σ' refers to the material under consideration and is either '*bf*' or '*p*' representing the base fluid and particles respectively. Thereby, essentially each of Eqn. 18 and 19 are composed of two equations, one or the fluid and the other for the particles. 'τ' represents the relaxation time with subscripts '*mom*' and '*θ*' representing momentum and thermal conditions respectively whereas '*F*' represents the forces acting on the system. The instantaneous forces acting on the particulate phase is modeled in accordance to Eqn. 6, with the force function determined within the computation at each spatial node per time step. The force term for the fluid component is expressed as

$$F^{bf} = F_{buoy} + F_D + F_B \tag{20}$$

The spatial arrangement of the particle velocity vectors are a function of the lattice structure and in the present study, a D2Q9 lattice system has been utilized, which has been illustrated in Fig. 2(b). The discrete particle velocities for such a lattice is mathematically described as

$$e_i = \begin{cases} e_0 = (0,0) \\ e_i = e(\cos\theta_i, \sin\theta_i), \theta_i = (i-1)\frac{\pi}{2}, \forall i = 1,2,3,4 \\ e_i = e\sqrt{2}(\cos\theta_i, \sin\theta_i), \theta_i = \frac{\pi}{4} + (i-5)\frac{\pi}{2}, \forall i = 5,6,7,8 \end{cases} \tag{21}$$

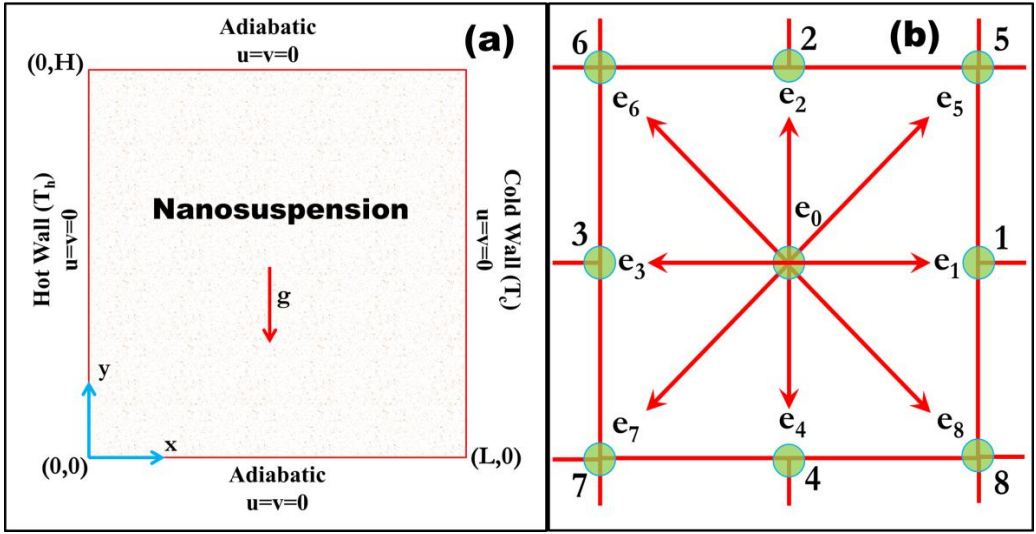

**Figure 2: (a)** Schematic of the computational domain and **(b)** the spatial arrangement of the velocity vectors for the D2Q9 lattice system used in the present case.

The hydrodynamic or momentum and thermal relaxation time for the base fluid and nanosuspension phases can be deduced from the ensemble averaged mathematical formulations for the kinematic viscosity and the thermal diffusivity of the fluids as

$$v^{bf} = \frac{2\tau^{\sigma}_{mom} - 1}{6} c^2 \Delta t \tag{22}$$

$$v^{ns} = \frac{2\sum_{\sigma} \bar{\rho}^{\sigma} \tau^{\sigma}_{mom} - 1}{6} c^2 \Delta t \tag{23}$$

$$\alpha^{ns} = \frac{2\sum_{\sigma} \bar{\rho}^{\sigma} \tau^{\sigma}_{\theta} - 1}{6} c^2 \Delta t \tag{24}$$

In the preceding equations, the mass fraction of the $\sigma^{th}$ component is expressed as

$$\bar{\rho}^{\sigma} = \rho^{\sigma} / \sum_{\sigma} \rho^{\sigma} \tag{25}$$

The LBM methodology requires defining the equilibrium distribution functions for the components such that all perturbed hydrodynamic and thermal fluctuations can be comprehended in terms of a datum energy function. The hydrodynamic and thermal equilibrium distribution functions are mathematically expressible in terms of the fluid velocity function and the lattice transport vectors as

$$f_i^{\sigma,eq} = W_i \rho^\sigma \left[ 1 + \frac{e_i . u^{\sigma,eq}}{c_s^2} + \frac{(e_i . u^{\sigma,eq})^2}{2c_s^4} - \frac{u^{\sigma,eq} . u^{\sigma,eq}}{2c_s^2} \right] \qquad (26)$$

$$g_i^{\sigma,eq} = W_i \theta^\sigma \left[ 1 + \frac{e_i . u^{\sigma,eq}}{c_s^2} + \frac{(e_i . u^{\sigma,eq})^2}{2c_s^4} - \frac{u^{\sigma,eq} . u^{\sigma,eq}}{2c_s^2} \right] \qquad (27)$$

The variable '$W$' denotes the mathematical weighing factor and its magnitudes are described as

$$W_i = \left\{ \left( \frac{4}{9}, i=0 \right); \left( \frac{1}{9}, i=1,2,3,4 \right); \left( \frac{1}{36}, i=5,6,7,8 \right) \right\} \qquad (28)$$

The instantaneous perturbed hydrodynamic and thermal distribution functions for the system are used to determine the instantaneous macroscopic properties of both the components as described by the following mathematical equations

$$\rho^\sigma(x,t) = \sum_i f_i^\sigma(x,t) \qquad (29)$$

$$\rho^\sigma(x,t) u^\sigma(x,t) = \sum_i f_i^\sigma(x,t) e_i \qquad (30)$$

$$\theta^\sigma(x,t) = \sum_i g_i^\sigma(x,t) \qquad (31)$$

The equilibrium velocity for the nanosuspension requires to be modeled based on the conservation of the effective localized mass flux of the particulate and the fluidic phases per unit area of the defined lattice structure per unit direction [33]. On similar grounds, the mean temperature of the suspension can be modeled based on the conservation of localized enthalpy of the particulate and the fluidic phases [34]. Mathematically, the two components can be expressed as

$$u^{ns,eq} = \frac{\{(\rho u)^{f,eq}(1-\varphi)\} + \{(\rho u)^{p,eq} \varphi\}}{\rho^f (1-\varphi) + \rho^p \varphi} \qquad (32)$$

$$\theta^{ns} = \frac{\{(\rho C_p \theta)^f (1-\varphi)\} + \{(\rho C_p \theta)^p \varphi\}}{\rho^f (1-\varphi) + \rho^p \varphi} \qquad (33)$$

The set of governing equations described in Eqn. (18) – (33) are used to solve natural convective thermal transport in alumina–water nanosuspensions in a square cavity domain as illustrated in Fig. 2(a). The domain consists of adiabatic top and bottom walls with isothermal hot and cold left and right vertical walls respectively. Zero concentration and no-slip conditions are imposed on each wall with the suspension within the domain considered as Newtonian and the mean flow to be laminar, incompressible and in accordance to the Boussinesq approximated body force. Mathematically, the no-slip boundary condition is imposed for the LBM in terms of a simplistic bounce-back scheme, wherein, the unknown discrete velocity momentum distribution function at any boundary lattice point is exactly equal in magnitude to the discrete velocity momentum distribution of the exact opposite node and directed along the former's image vector when reflected against the wall under consideration. Mathematically, it is expressible as

$$f(x=0)_{e_i} = f(x=0)_{-e_i}, f(x=L)_{e_i} = f(x=L)_{-e_i} \tag{34a}$$

$$f(y=0)_{e_i} = f(y=0)_{-e_i}, f(y=H)_{e_i} = f(y=H)_{-e_i} \tag{34b}$$

The isothermal boundaries are mathematically imposed via conservation of the thermal flux across the boundaries and expressed in Eqn. 35(a) and (b), whereas the adiabatic condition is imposed by reducing the thermal flux to null via finite difference approach as illustrated in Eqn. 36.

$$\frac{g(x=0)_{e_i} + g(x=0)_{-e_i}}{W_{e_i} + W_{-e_i}} = T_h \tag{35a}$$

$$\frac{g(x=L)_{e_i} + g(x=L)_{-e_i}}{W_{e_i} + W_{-e_i}} = T_c \tag{35b}$$

$$g(y=0)_{e_i} = g(y=1)_{-e_i}, f(y=H)_{e_i} = f(y=H-1)_{-e_i} \tag{36}$$

The LBM formulation in the present study has been coded in FORTRAN and simulations have been performed using the in-house developed code to solve natural convective transport in the square cavity as discussed above. Simulations have been performed for the Ra range of $7 \times 10^5$ to 3.39 x $10^6$ and Pr = 8.923 in order to validate the results with experimental observations [14]. Lattice independence study in order to determine the computational time and accuracy optimized lattice structure is performed for Ra = $10^6$ for 4 % alumina (33 nm)-water nanosuspensions considering all the slip forces as detailed in Table 1. The lattice structure has been represented as the grid parameter 'n', such that it corresponds to 'n x n' lattice system. The average Nusselt number at the hot wall with respect to the grid parameter and the error suffered with respect to change in grid parameter has been illustrated in Fig. 3(a). Based on analysis, the '160 x 160' lattice has been chosen since it poses an optimality when both accuracy and computational intensiveness are considered as major criteria. The code has been validated against published numerical solutions for air [35, 36] and water [37] and the same has been illustrated in Fig. 3(b), wherein good accuracy has been observed, thereby establishing the credibility of the present formulation.

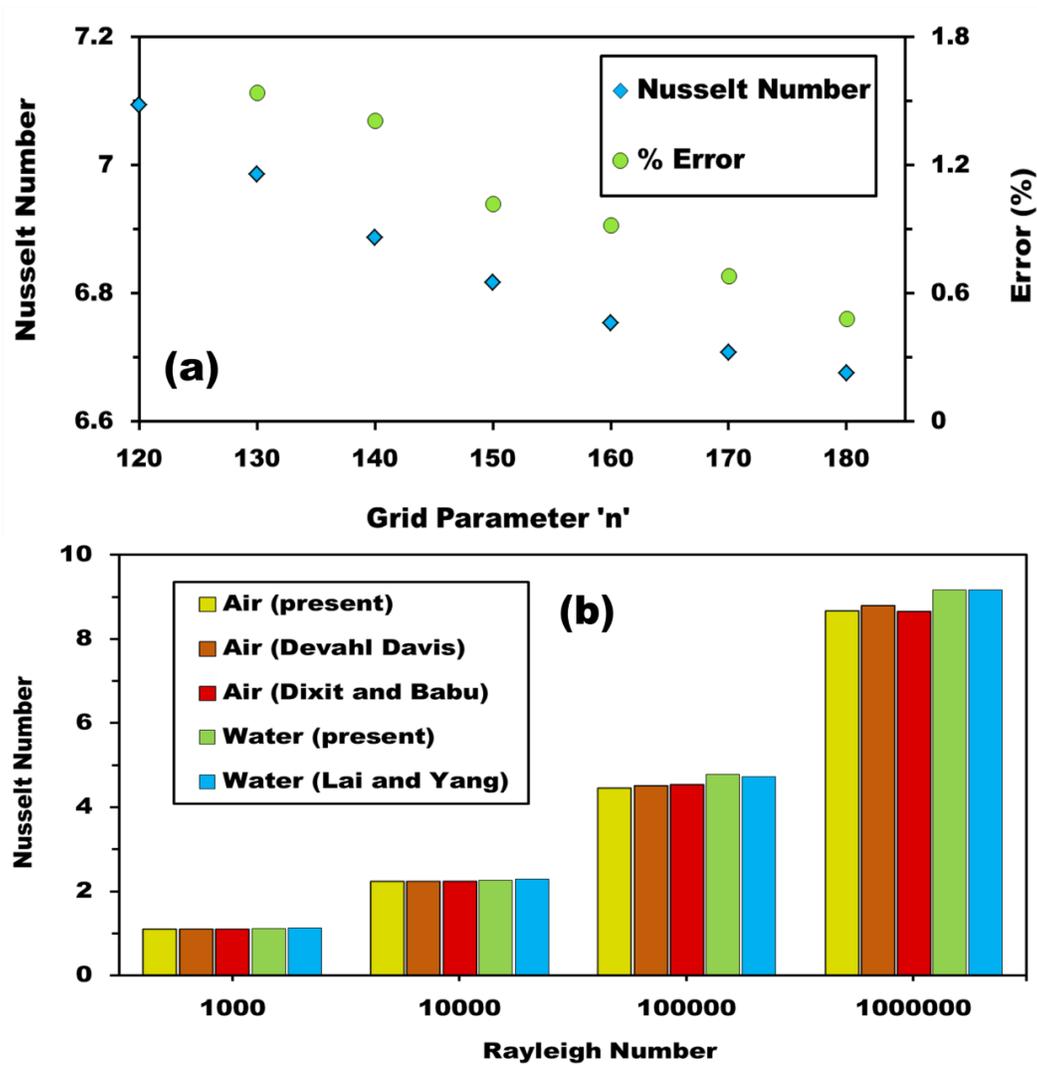

**Figure 3: (a)** Error encountered within the simulation with respect to the chosen grid parameter and **(b)** validation of the present MCLBM formulation with published reports (the cases presented in the plot for a given Ra, from left to right exactly correspond to those in the legend, from top to bottom).

## 3. Results and Discussions

In an attempt to comprehend the physics of particle-fluid interactivities and its impact vis-à-vis buoyancy driven thermal transport in nanosuspension systems, it is necessary to meticulously analyze the contributions and mechanisms behind each participating slip force. In case of the present problem at hand, simulations have been carried out initially considering all the slip forces in action. Further, all the slip forces have been eliminated one by one to understand the mechanism and degree to which each force contributes to the deterioration of thermal transport. Natural convective heat transport in nanosuspensions has been a topic of debate over the past two decades. It arises from the fact that while deterioration is observed in case of experimentations, the predictions via conventional numerical simulations predict enhanced transport. Few non-conventional approaches to model such systems have led to observed deterioration, however, only qualitatively. As per the knowledge of the authors, this is the first ever report of a numerical scheme which can accurately predict the deterioration, both quantitatively and qualitatively. The present simulations have been initially performed considering all the slip mechanisms in action, so as to predict the deteriorated thermal performance. Then, each slip force is kept dormant while others active so as to understand the degree and mechanisms by which each slip deteriorated heat transfer. Finally, the effects of governing parameters such as Rayleigh number, particle size and material etc. have been exhaustively studied to provide a clear picture of natural convection thermal transport in colloidal nanosuspensions. The predictive performance of the numerical formulation against experimental data has been illustrated in Fig. 4.

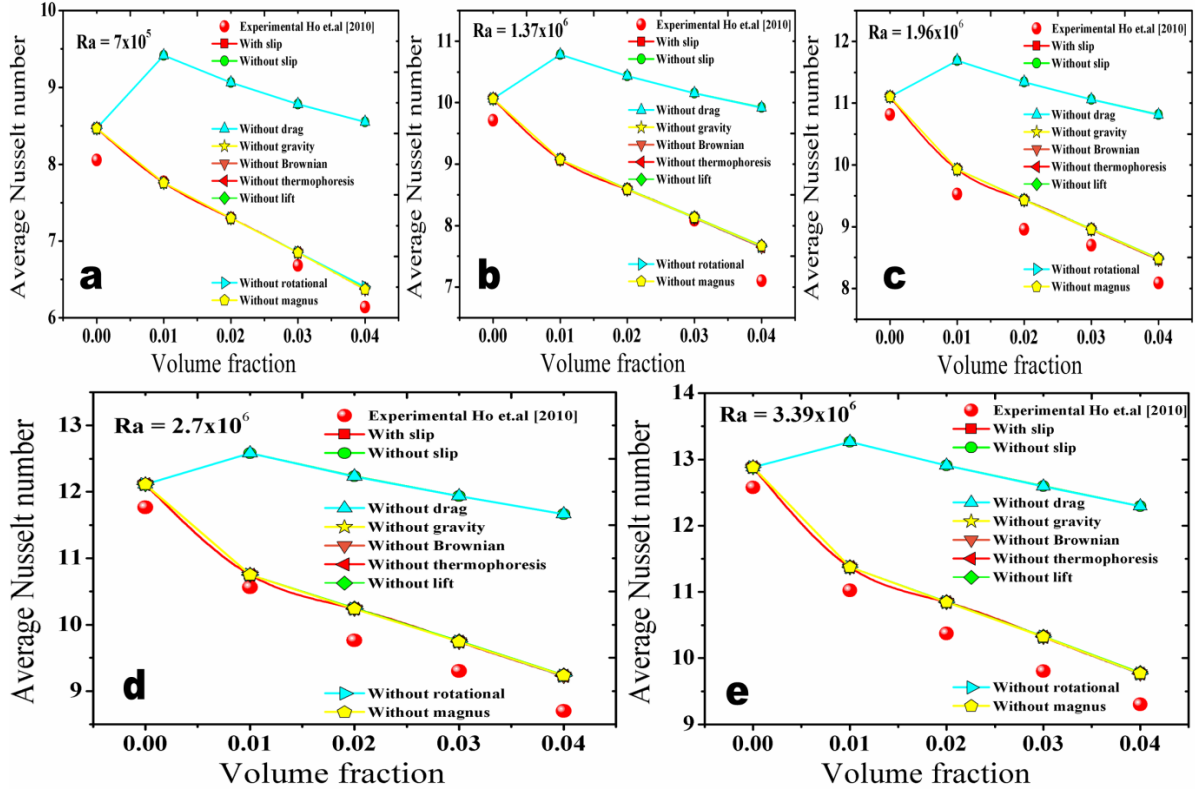

**Figure 4:** Validation of the results for heat transport in nanosuspensions, as obtained from the MCLBM formulation against experimental reports [14].

As obvious from Fig. 4, the present formulation accurately predicts the deterioration in thermal transport, thereby cementing the hypothesis of the importance of the slip mechanisms in convective thermal transport in colloidal nanosuspensions. Analysis of results suggests that the drag slip has the maximal effect in deteriorating heat transfer as compared to all other slip forces combined. As a matter of fact, when the drag force is switched off in the formulation, the trend predicted by the simulation nearly matches with the cases wherein none of the slip mechanisms have been considered at all. Further studies show that as far as degree of reduction in transport in concerned, drag is followed by Brownian and the thermophoretic slips. This proves conclusively that unlike forced convection, wherein drag is not the leading slip component, the physics of thermal transport in natural convection of colloids is completely different. The force of drag arises due to the difference in velocities of the fluid molecules and the particles, who fail to cope with the fluid due to higher retarding inertia caused by gravity. The force of drag is directly proportional to the difference and further retards the particle velocity, which in turn reduces the velocity of the fluid molecules in its immediate vicinity, leading to deteriorated transport of heat. Heat transfer is observed to deteriorate further with increasing concentration of nanoparticles in the system. Increased particles in the system adds increases the localized effective drag of fluid molecules in the immediate vicinity of the particles, thereby decreasing the thermal transport capabilities. Natural convective thermal transport depends on a plethora of governing parameters, both material and process. In order to clearly comprehend the effects of particle migration in deterioration of transport, a detailed parameterization of the study is required. Based on analysis of such data, detailed discussions on the effect of controlling parameters and implications vis-à-vis particle migration and interactions have been provided in subsequent sections.

## 3. a. Strength of natural convection

The dominance of the drag slip in the present case has already been established in the preceding section, however, the degree of dominance and the implications require to be understood. The effect of strength of natural convection, governed by Ra, on the heat transfer deterioration at various concentrations has been illustrated in Fig 5. The effect of slip has been clearly demonstrated and its absence can be seen to predict enhancement in thermal transport rather than deterioration, a result which has often been reported in numerical works pertaining to the present topic. Interestingly, the case of the drag slip being solely switched off to inactivity corresponds closely to that of the case of no slip, thereby cementing the dominating role of drag slip in such processes. Interestingly however, at lower Ra ranges of $10^3$–$10^4$, the formulation predicts that it is actually possible to achieve enhanced thermal transport in natural convection of nanosuspensions. At low Ra~$10^3$, the addition of a minute amount of nanoparticles (~ 1 wt. %) to the fluid yields highly augmented heat transfer, which then reduces with increasing concentration. however, the reduced value remains greater in magnitude than that for the pure fluid, to concentrations as high as 4 wt. % and beyond. A similar trend is observed in moderate Ra ~ $10^4$, wherein a small enhancement is observed with initial seeding, but the transport deteriorates once the concentration becomes ~ 1 wt. %. It is at this Ra that the strength of the present formulation is observed, such that, the absence of slip forces predict augmented transport at any given concentration, whereas that considering slip (predominated by drag) predicts deteriorated thermal transport.

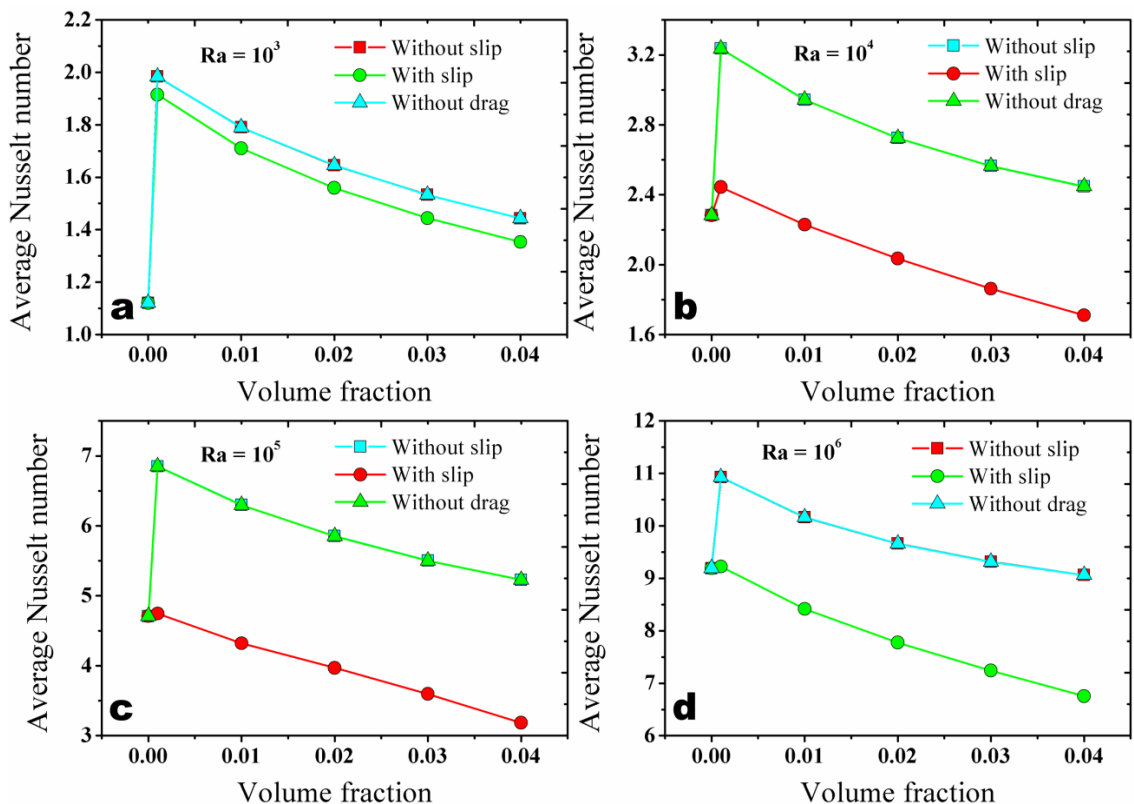

**Figure 5:** Effects of Rayleigh number and the incorporation of slip mechanisms on the deteriorated heat transfer performance of nanosuspensions

Although high accuracy is observed as far as predictability of thermal transport is concerned, the mechanics behind Ra dependent deterioration should be probed. Ra governs the mode of thermal transport in a natural convective system, wherein the predominant mode is conduction when the Ra is lesser than the critical value for the system and convection when it is greater than the critical value. For natural convection in enclosed differentially heated cavities, the critical Ra is often close to the magnitude of $\sim 10^3$ [ref]. Along such lines, one would expect deteriorated thermal transport even in the case of Ra $\sim 10^3$, since the presence of nanoparticles have been reported to deteriorate natural convection heat transfer. However, interestingly, it is not so and the formulation predicts enhanced transport at lower Ra values, where once would expect onset of convective transport. This can be explained based on findings that nanoparticles in the colloidal state also act as deterrents to the onset of convective transport [38], which is often portrayed by a larger magnitude of the critical Ra than that expected from the same system operating with pure fluids. Convective transport begins when the buoyant forces strongly overcome the momentum diffusivity of the fluidic system. The presence of nanoparticles in the system augments the momentum diffusivity of the fluid [30], thereby delaying the onset of buoyancy driven convection. The presence of nanoparticles of materials with higher thermal conductivities compared to their heat capacities also leads to augmented thermal diffusivity of the fluid, thereby leading to further delay of onset of thermally governed buoyancy driven natural convection in such colloids. Therefore, the enhanced thermal transport is observed al low Ra solely due to the delayed onset of convection and predominant transport by conduction. Numerous studies report enhanced thermal conductivity of nanosuspensions over their base fluids [3, 4]. In the conductive regime, the transport of heat from the hot to the cold side is dominated by conduction by the fluid molecules, thermophoretic diffusion of the nanoparticles extracting heat from the hot wall followed by Brownian diffusion of the particles, rejecting additional heat into the fluidic medium. The presence of the nanoparticles enhances the effective thermal conduction capabilities of the suspension, leading to its manifestation as augmented Nu of the system compared to the base fluid. However, the crux of the study reveals that once in the natural convective regime, the dominance of drag force intervenes, leading to deteriorated transport of heat compared to the base fluid.

## 3. b. Size of dispersed nanoparticles

The role of the particles in deterioration of heat transfer only surfaces once the combined effects of thermal and momentum diffusivity is overshadowed by the buoyant forces, leading to established natural convective transport. As soon as components of velocity are imparted to the fluid molecules due to density gradient driven convective cells, momentum slip between the particles and the fluid molecules arise, leading to the various slip mechanisms. The slip mechanisms play a crucial role in modifying the thermal transport characteristics of the system apart from its performance due to augmented thermophysical properties. In order to understand the effects of the particulate phase in depth, light needs to be shed on the role of particle phase material, size and concentration in deterioration of the transport phenomenon. In nanosuspensions, physical properties have been reported to be strong functions of particulate size [11]. Similarly, the degree of deterioration of transport is a strong function of dispersed phase particulate size. The effect of particle size on the convective thermal transport has been illustrated in Fig. 6.

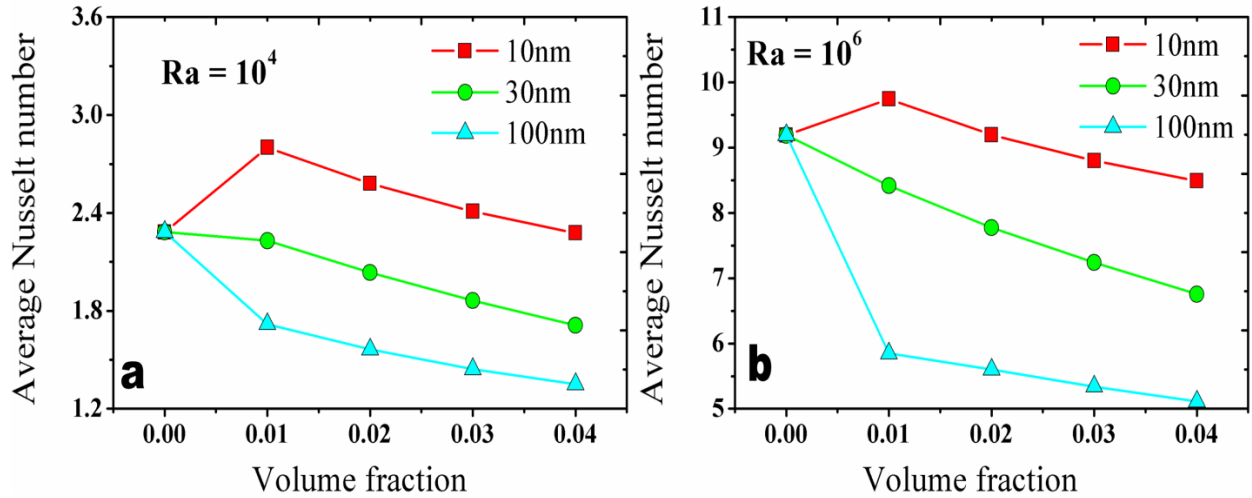

**Figure 6:** Effect of particle size on natural convective heat transfer characteristics in nanosuspensions at different Ra.

The computations reveal that thermal transport by natural convection in nanosuspensions is strongly influenced by the average size of the dispersed particulate phase. As observable from Fig. 6, the deterioration of thermal transport follows an inversely proportional trend with respect to the particle size. For aluminum oxide nanoparticles of the order of 30 nm and above, deterioration is observed at all volume fractions and for both low and high governing Ra. However, interestingly, deteriorated transport is not always true for natural convective thermal transport in nanosuspensions. As observable, Nu is enhanced at all Ra ranges for dilute suspensions (~ 0.01 volume fraction) of ~ 10 nm particles. It is noteworthy that the enhancement in transport is much higher at lower Ra (~ 20%) compared to the higher Ra condition (~ 5%). Beyond the dilute regime, the transport deteriorates, but does not deteriorate compared to the base fluid for concentrations as high as 0.04 volume fraction for low Ra. For the high Ra condition, the deterioration compared to base fluid occurs around ~ 0.02 volume fraction. Such observations reveal many important phenomena. At lower Ra, inception of convective transport against buoyancy occurs and the convective currents are weak in strength. As a result, partial conductive transport still dominates specific regions of the domain wherein the buoyancy driven transport is yet to transform to the dominant form. In case of smallest particle diameters, the Brownian and thermophoretic fluxes are maximal and the drag force is minimal. At low Ra, wherein the convective transport is weak enough to pose negligible damping of the slip mechanisms, high heat conduction across the domain exists, leading to enhanced Nu, which is further augmented at lower concentrations due to least enhancement of viscous forces within the suspension. As the concentration increases, the viscosity of the suspension increases due to the presence of the particles, which in turn reduces the degree of enhancement observed. At higher Ra, convective transport dominates, leading to increased drag force, which even though minimal for small particles, reduces the degree of enhancement in transport caused by Brownian and thermophoretic flux. This is justified from the fact that at 1 wt. % and low Ra, the enhancement in Nu observed is ~ 33%, which in the case of high Ra reduces to ~ 5.4 %. Furthermore, as the particle size increases, the Brownian and thermophoretic fluxes reduces considerably whereas the drag forces increase, leading to higher dissipation of particle linear momentum within the system. This leads to disruption of the convective currents within smaller length scales compared to that of the base fluid, causing reduction in heat transport efficiency. With increasing particle size, concentration and Ra, the deterioration in transport enhances due to greatly increased drag and interparticle collisions, which hamper the convective flux. The effect of dominance of drag is established from the effect of Ra. For a system with dispersed particle size of 100 nm at 4 wt. % concentration, the deterioration in Nu enhances from ~76 % to 85 % with increase of Ra from conductive regime to purely convective regime. With increasing Ra, the buoyant convective transport increases, leading to increase in fluidic velocity, which leads to enhanced drag caused by increased relative velocity between the fluid and the particles. Thereby, the importance of the drag slip on deteriorative effect on thermal transport is established through theoretical discussions.

### 3. c. Concentration of nanoparticles

The transport is governed by multiple parameters, among which the concentration of the dispersed phase plays a crucial role in determining the degree of deterioration in heat transfer. Concentration of nanosuspensions plays a major role in designing cooling fluids for thermo–fluidic systems since enhanced concentrations have been reported to yield higher thermal conductivities in such suspensions. However, in systems where natural circulation cooling is exploited, an efficient compromise between the enhancement in thermal conductivity and the deterioration in natural convective transport has to be obtained, since both are strong functions of the suspension concentration. The effect of particle concentration on the heat transfer capabilities of the nanosuspension has been illustrated in Fig. 7.

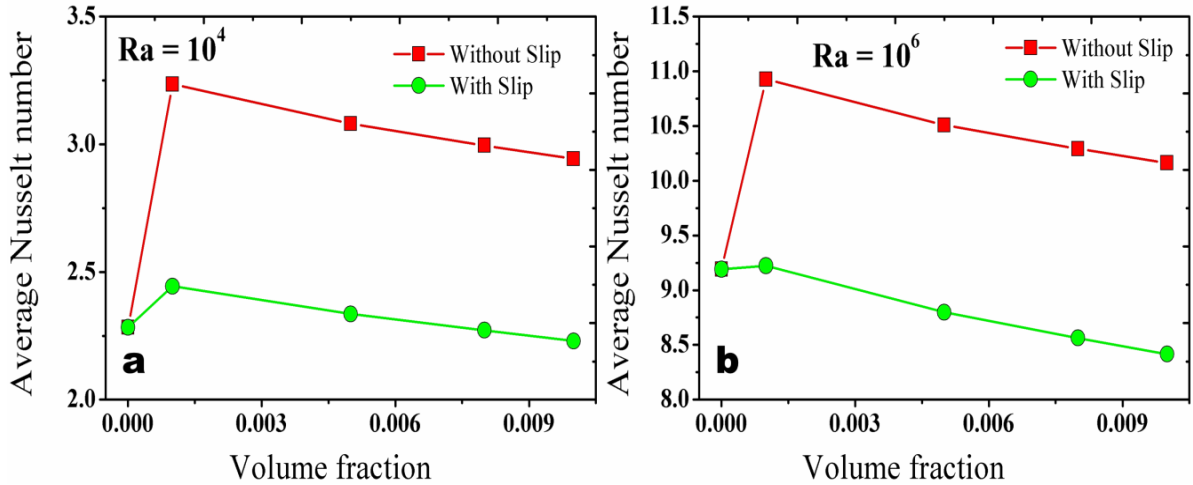

**Figure 7:** Effect of nanoparticle concentration and the importance of slip forces in predicting the degree of deterioration in nanosuspensions.

The heat transfer capability of the suspension can be observed to deteriorate at all Ra, ranging from low to high, when all the slip mechanisms; especially drag, are considered. In the absence of the slip forces, enhanced transport is observed. The absence of slip leads to sudden augmentation in heat transfer due to the presence of nanoparticles, similar to the case of forced convection. With increase in concentration, the effective viscosity utilized in the computation enhances, thereby reducing the Ra, which leads to further reduction of the Nu in accordance to the dependence of Nu on Ra for natural convection. Since the effective property models fail to capture the enhancement of viscosity by large margins, the degradation in Ra is low, leading to further reduced deterioration in heat transport. This in essence further establishes the fact that non–incorporation of the slip forces in the numerical model leads to predictions that counter observations. On the contrary, when the slip forces are incorporated within the formulation, the computations yield deteriorated transport when compared to the base fluid itself. At low Ra, wherein the conductive dominance still persists, lower values of concentration lead to low suspension viscosities. The low values of viscosity (often comparable to the viscosity of the base fluid in magnitude) conjugated with the conductive dominance leads to slightly augmented heat transport characteristics. However, owing to the drag force, the traversal of the particles is severely prohibited within the fluid medium, leading to much lower enhancements as compared to the case wherein slip mechanisms are not incorporated. Also, with increasing concentration, the stability of the suspension [5] reduces due to increased particle–particle interaction mediated agglomeration and de–agglomeration dynamics, which leads to deteriorated thermal transport in real systems. Equivalently, the same mechanism is reflected as enhanced particle–particle collisions within the numerical scheme, leading to reduced magnitude of localized particle velocity, which leads to reduced transport of heat due to particle migration. Enhanced viscosity and collision mediated reduction in particle migration, thereby reducing thermal transport by the fluid and particle migration respectively are jointly responsible for deterioration in natural convective heat transport with increasing concentration.

The major strength of the proposed Multi–Component LBM (MCLBM) lies in the fact that it can efficiently address the issue of particle migration effects by simultaneous yet independent modeling of the fluid and the suspended phases. Models of lower complexities, such as Single Component Non–Homogeneous LBM (SCNHLBM) have also been proposed, wherein the nanosuspension is treated as a single component fluid, however, with induced non–homogeneity so as to capture the apparent gradients in concentration which are created due to the convective motion. Such models have been found to assess the trend of deteriorated transport qualitatively; however, they fail to qualify the deterioration. Although such models can focus on the importance of slip mechanism in such systems, quantification of deterioration is not possible since the most dominant transport deteriorating slip mechanism of drag cannot be modeled in a single phase system. Thereby, incorporation of drag forces and simulating the species transport within the system due to concentration gradients generated is the crux strength of the present model. Computationally derived contours of concentration of the suspension at steady state have been illustrated in Fig. 8 (a)–(d). Interestingly, clearly distinguishable concentration contours are obtained from the MCLBM formulation, which indicates that migration of population of the dispersed phase and redistribution thereof is an important parameter that governs heat transfer in such systems.

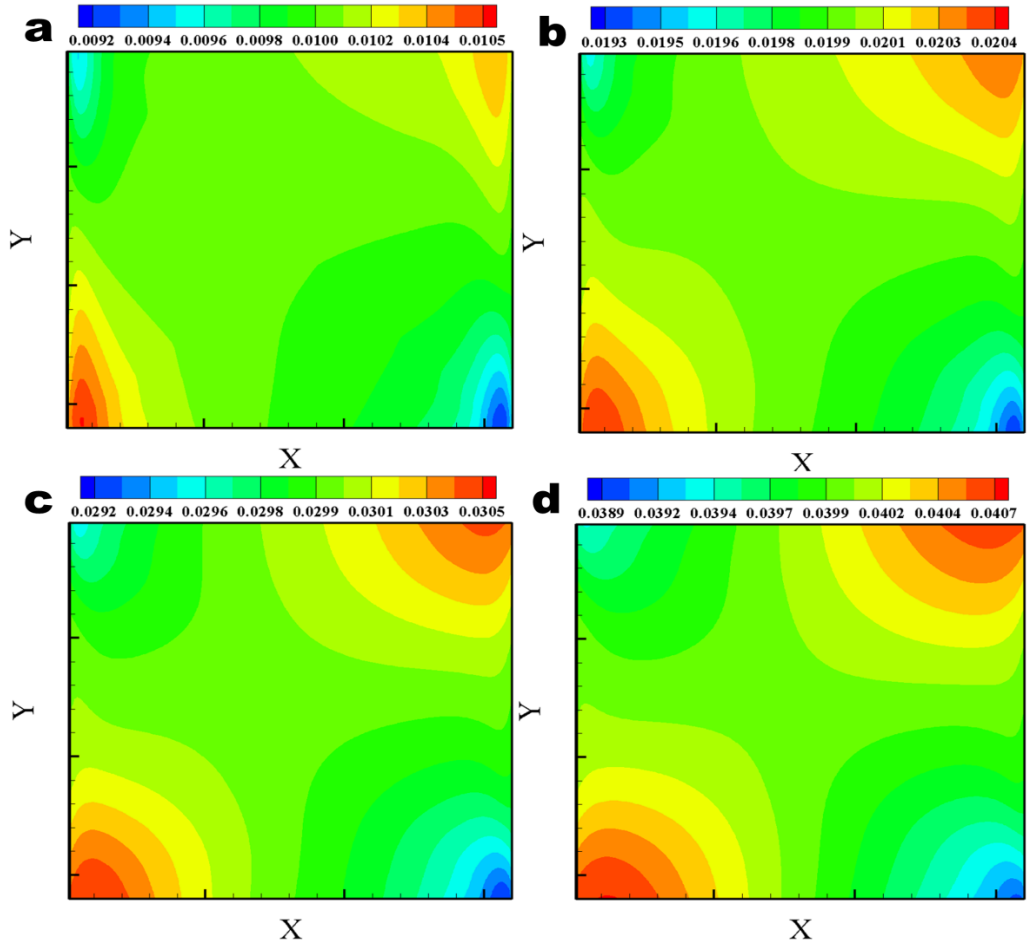

**Figure 8:** Contours of concentration within the domain at steady state for different concentrations **(a)** 1 wt. % **(b)** 2 wt. % **(c)** 3 wt. % **(d)** 4 wt. % of the nanosuspension.

The concentration contours are observed to be dependent on the initial concentration of the nanosuspension, which is intuitively correct since the number density of nanoparticles are expected to govern the concentration distribution once the system reaches steady state. As observable, the circulation pattern of the fluid leads to augmented concentration at the lower regions of the vertical hot wall and the upper region of the vertical cold walls, whereas, reduced concentration is observed at the bottom region of the vertical cold wall. Furthermore, as observable from the contours, the fluid containing the initial concentration of the suspension is confined to the central region of the cavity and the confinement ratio (defined crudely as the ratio of area within which the fluid with initial concentration is confined to, to the total area of the domain) decreases with increasing initial concentration of the suspension. Also, if the ratio of the maximum to the minimum concentration within the cavity is considered, the magnitude is maximum (1.141) at 1 wt. % and reduces with increasing concentration, 1.057 at 2 wt. % and decays to 1.045 for both 3 and 4 wt. %. As the concentration increases near the bottom of the hot wall, the viscosity of the local fluid increases, thereby decreasing the local Ra, which in turn reduces the Nu corresponding to thermal extraction by the fluid at the hot wall. The accumulation and stagnation of particles also leads to increased localized relative velocity at the region, leading to enhanced drag on the particles and fluid molecules passing close to the region of accumulation. At the base of the hot wall, this essentially hampers the free circulation of fluid from the bottom of the enclosure towards the top, leading to further reduction in heat extraction efficiency. With increasing concentration, the viscous drag within the system increases and the effective size of the zone of accumulation also enhances, the coupled effect thereby further reducing thermal extraction from hot wall.

The accumulation of particles at the top end of the cold wall leads to enhanced drag on the fluid elements and particles traversing from the top towards the bottom of the cold wall, thereby further reducing deposition of heat on to the cold wall. Furthermore, enhanced concentration at the top of the wall leads to reduced stability, thereby leading to further reduction in heat sink effect of the cold wall. The zone of depleted concentration at the bottom of the cold wall also leads to reduction in the local Nu between the fluid and the wall due to decreased heat transfer by particle migration effect. Also, majority of the domain experiences the presence of severely diluted form of the nanosuspension compared to its initial concentration, resulting in lowered thermal transport by particle migration effects. The size of the zone of depleted concentration also increases in magnitude with increasing initial concentration, thereby leading to enhanced deterioration for concentrated nanosuspensions. In essence, the combined effects of reduced particle migration effects due to reduced concentration profiles and deteriorated heat extraction and deposition due to increased viscosity and reduced stability near the thermally significant walls lead to deteriorated transport compared to the base fluid.

## 3. d. Physical properties of fluid and nanoparticles

The effective influence of the slip mechanisms on the transport to heat by natural convection in a nanosuspension is a strong function of the physical properties of the base fluid and the nanoparticle material. This is inevitable since the magnitude of the slip forces are dependent on the viscous resistance offered by the fluid and the stability of the suspension is governed by the differences in densities of the dispersed and the fluid medium. The effect of base fluid on the deterioration of transport has been illustrated in Fig. 9. As observed, at both low and high Ra and low concentration of particles, the deterioration is much more pronounced in case of Ethylene Glycol (EG) and EG–water (1:1) mixtures than that of pure water. This conclusively provides evidence on the dominance of drag slip on the deterioration of transport. Given the higher viscosities of water-EG and EG over water, the particles experience enhanced degrees of retardation due to viscous damping of motion, thereby leading to higher drag force on the fluid elements, leading to reduced heat transfer. However, with increasing concentration of nanoparticles, the effective deterioration in case of water becomes nearly equivalent to that of water-EG mixture or EG. Such a phenomenon is observed due to the reduced stability of the suspension at higher concentrations when a low viscosity fluid is under question. The stability of suspensions of given concentration is more in case of water-EG or EG than compared to water in accordance to Stokesian mechanics of particulate material suspended in fluid medium. Accordingly, water based suspensions of high concentrations experience extra deterioration in transport owing to the reduction in heat transfer due to erratic concentration distribution (as explained within the preceding subsection) within the fluid medium as compared to its viscous counterparts. Essentially, the fact that the degraded transport is governed by the coupled mechanisms of imbalanced concentration distribution and viscous drag is further established.

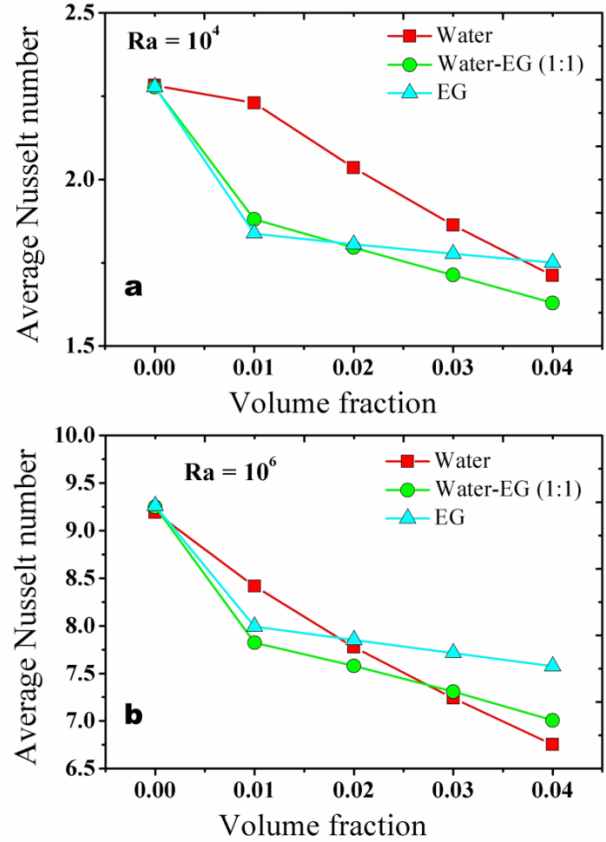

**Figure 9:** Effect of base fluid thermophysical properties, especially viscosity, on the extent to which natural convective thermal transport suffers deterioration in nanosuspensions.

In similitude to the thermophysical properties of the base fluid strongly governing the transport phenomena, the thermophysical properties of the nanoparticle material also governs the extent to which the transport is deteriorated. Such an observation essentially hints towards the fact that stability of the nanosuspension is also an important factor as far as thermal transport is concerned. The deterioration of transport for various nanosuspensions as compared to water has been illustrated in Fig. 10. Particles of materials of higher densities tend to possess higher magnitudes of settling velocity in accordance to the expression of Stokesian terminal velocity as compared to particles of low densities. With increasing propensity to gravitational effects, heavier particles therefore tend to increase the concentration towards the lower part of the domain, which in turn leads to enhanced fluidic drag within that region as well as the suspension is depraved of the requisite amount of particles for effective heat extraction or deposition from or to the walls in the remaining regions of the domain. As a consequence, the heat transfer is further deteriorated over and above the deterioration due to the normal concentration non–uniformity. As evident from Fig. 10, particles of low density, such as $SiO_2$ in fact leads to enhanced transport as compared to water due to highly augmented heat transfer by particle migration, which supersedes the viscous drag and concentration deficit effects. However, for particles with higher densities, viz. $Al_2O_3$, CuO and Cu, the deterioration is inevitable, the deterioration increasing with increasing concentration, which has also been reported from experiments [11]. It is interesting to note that higher thermal conductivity has a much reduced role in governing convective transport as compared to the particle density, as evident from the fact that copper (with a higher density and thermal conductivity) essentially deteriorates transport much more than alumina (with lower density and thermal conductivity). Similarly, copper oxide and alumina have relatable magnitudes of thermal conductivity, however, owing to larger density; the transport is deteriorated more in the former's case.

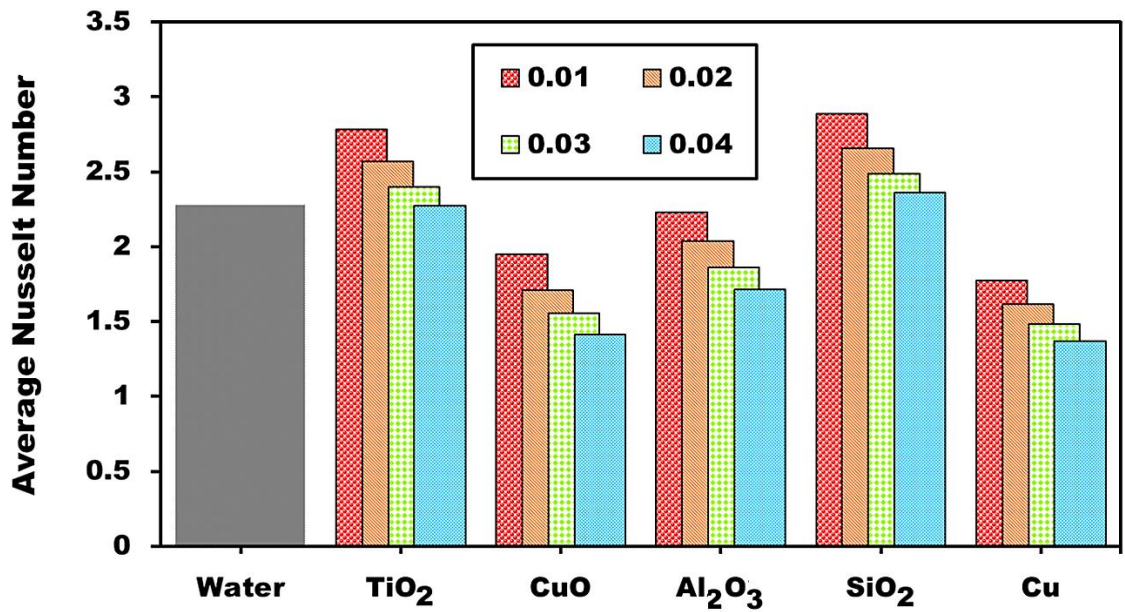

**Figure 10:** Effect of nanoparticle (of same sizes) material and concentrations on the heat transfer deterioration in nanosuspensions. The legend refers to the suspension concentration (wt. /wt.)

### 3. e. Aspect ratio of the convective system

Flow pattern and structure has also been observed to contribute to the extent of deterioration of transport. The heat transfer capabilities of the system and the associated streamlines and isotherms for different aspect ratios have been illustrated in Fig. 11 and 12.

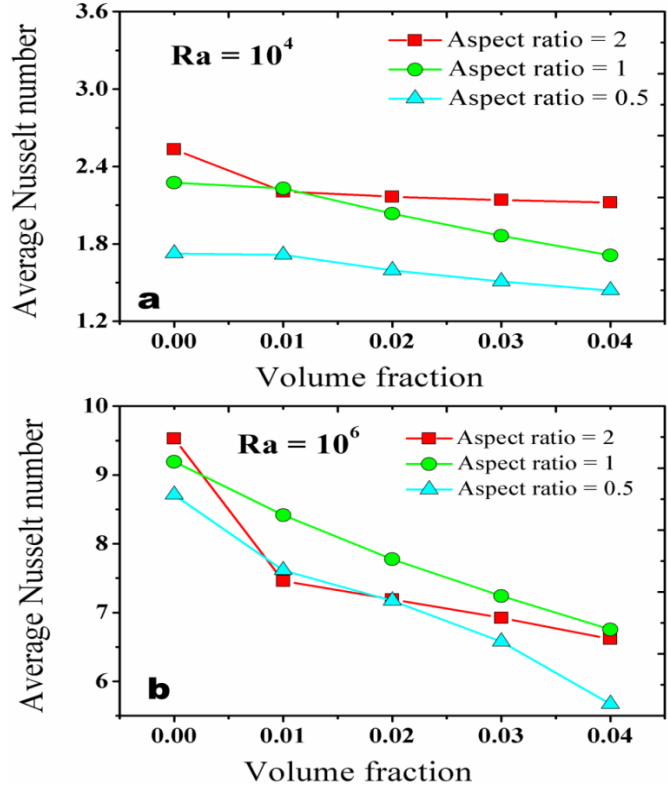

**Figure 11:** Effect of domain aspect ratio and Ra on degradation of heat transfer during natural convection in nanosuspensions.

At low Ra, conductive transport still predominates, and as a result, the effect of aspect ratio on heat transfer is negligible, as observable in Fig. 11(a). Transport of heat is mostly via conduction, and as such the geometry of the domain poses very little to the deterioration process, which leads to fairly constant heat transfer with respect to concentration. The differences observed for different aspect ratios are solely due differences in surface area available for heat extraction and deposition. However, at high Ra, the effect of geometry as well as concentration dually leads to highly deteriorated transport. However, there exist marked differences between the heat transfer characteristics for aspect rations 2 and 0.5. Such observations maybe explained based on the streamlines and isotherms related to the transport process. In case of aspect ratio 2, the cavity is horizontally constricted as compared to aspect ratio 1, whereas vertically twin fold. As a result, the suspended particles need to traverse a reduced horizontal distance, which severely reduces the time scale of exposure to viscous drag. Furthermore, the constricted space leads to more uniformity in the concentration distribution compared to aspect ratio of 1. Furthermore, the height of the cavity ensures that the zones of enhanced and depleted concentrations are far apart than that of aspect ratio 1, thereby ensuring more heat extraction or deposition by the suspension. Altogether, these factors improve heat transfer in aspect ratio 2 as compared 1, whereas the converse leads to further deteriorated transport for 0.5.

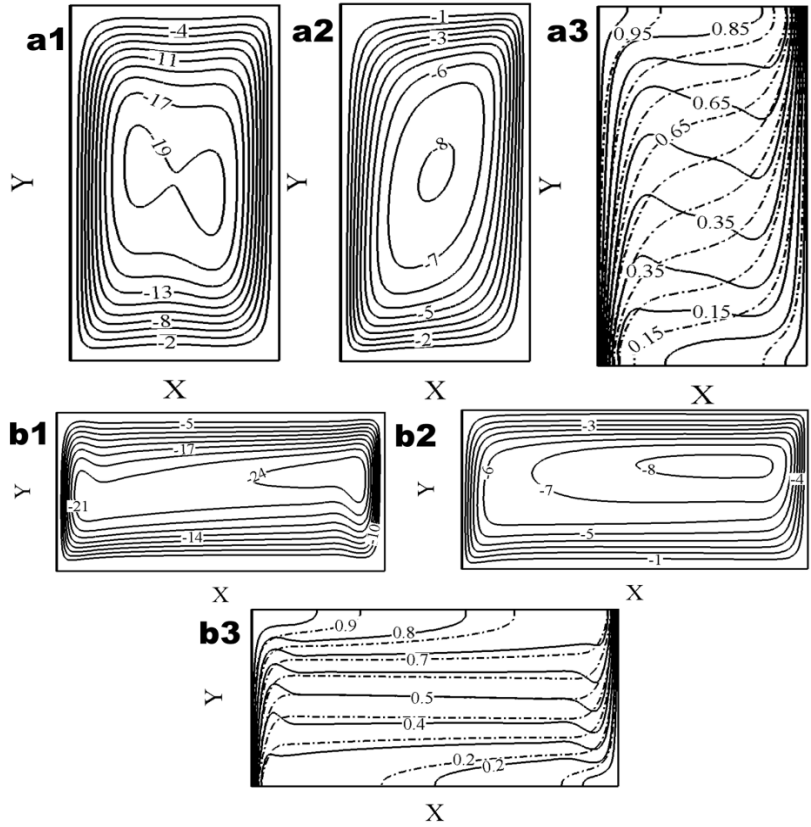

**Figure 12:** Line maps of **(a1 and b1)** streamlines for water **(a2 and b2)** streamlines for nanosuspensions and **(a3 and b3)** isotherms for water (chained) and nanosuspensions (continuous) for different aspect ratios of the computational domain.

## 4. Conclusions

The present work focuses on developing a multi component LBM formulation which can effectively track the variant interparticle and particle-fluid slip interactions and thereby accurately predict the heat transfer caliber of nanosuspension systems during natural convection. The article is the first of its kind to bridge the gap between experiments and simulations pertaining to natural convective thermal transport in nanosuspensions, wherein, till date, experimental and numerical simulations (by conventional formulations) traversed along different paths. A simplistic scaling analysis to comprehend the effects of slip mechanisms on natural convective heat transport yields that drag is the predominant slip mechanism and therefore, the major process governing entity for the observed deterioration in heat transfer. The multi component LBM segregates the particle and the fluid phases while in essence solving the two systems in a coupled fashion. The formulation has been used to simulate natural convection of nanosuspensions in a square cavity and the results have been observed to accurately track experimental findings; qualitatively and quantitatively. The present model is capable of predicting experimental results of deteriorated heat transfer in natural convection of nanosuspensions over the entire range of suspension concentration and system Ra. Furthermore, a detailed parametric study to reveal the effects of various isolated slip forces, Ra, concentration, particle size, thermophysical properties of the base fluid and particle material and aspect ratio of the convective domain on the heat transfer capabilities has been performed. The present study, in addition to predicting deteriorated natural convective heat transfer in nanosuspensions, also sheds insight onto the system physics and mechanisms which are responsible for the observed transport phenomena.

## Acknowledgements

The authors would like to acknowledge usage of the Virgo supercomputer cluster at the Computer Centre of the Indian Institute of Technology, Madras for executing the simulations. PD would also like to thank the Ministry of Human Resource and Development (MHRD), Govt. of India, for the doctoral research scholarship.